\documentclass[aps,prl,twocolumn,superscriptaddress,showpacs]{revtex4}
\usepackage[dvips]{graphicx}
\usepackage{bm}
\usepackage{dcolumn}

\def\lamb#1#2{$^{#1}_{\Lambda}${#2}} 
\def\lam#1#2{$^{#1}_{~\Lambda}${#2}} 

\begin{document}

\title {High Resolution Spectroscopy of $^{12}_{~\Lambda}$B by
Electroproduction}

\author{M.~Iodice}                                                   
\affiliation{Istituto Nazionale di Fisica Nucleare, Sezione di Roma Tre, 
I-00146 Roma, Italy}

\author{F.~Cusanno}                                                   
\affiliation{Istituto Nazionale di Fisica Nucleare, Sezione di Roma1, 
Piazza A. Moro, Rome, Italy}

\author{A.~Acha}
\affiliation{Florida International University, Miami, Florida 33199, USA}

\author{P.~Ambrozewicz}
\affiliation{Florida International University, Miami, Florida 33199, USA}

\author{K.A.~Aniol}
\affiliation{ California State University, Los Angeles, Los Angeles, 
California 90032, USA}

\author{P.~Baturin}
\affiliation{Rutgers, The State University of New Jersey, Piscataway, 
New Jersey 08855, USA}  

\author{P.Y.~Bertin}
\affiliation{Universit\'{e} Blaise Pascal/IN2P3, F-63177 Aubi\`{e}re, France}

\author{H.~Benaoum}
\affiliation{Syracuse University, Syracuse, New York 13244, USA} 

\author{K.I.~Blomqvist}
\affiliation{Universit\"at Mainz, Mainz, Germany}

\author{W.U.~Boeglin}
\affiliation{University of Maryland, College Park, Maryland 20742, USA}    

\author{H.~Breuer}
\affiliation{University of Maryland, College Park, Maryland 20742, USA}    

\author{P.~Brindza}
\affiliation{Thomas Jefferson National Accelerator Facility, Newport News, 
Virginia 23606, USA}

\author{P.~Byd\v{z}ovsk\'y}
\affiliation{Nuclear Physics Institute, \v{R}e\v{z} near Prague, Czech
Republic}

\author{A.~Camsonne}
\affiliation{Universit\'{e} Blaise Pascal/IN2P3, F-63177 Aubi\`{e}re, France}

\author{C.C.~Chang}
\affiliation{University of Maryland, College Park, Maryland 20742, USA}    

\author{J.-P.~Chen}
\affiliation{Thomas Jefferson National Accelerator Facility, Newport News, 
Virginia 23606, USA}

\author{Seonho~Choi}
\affiliation{Temple University, Philadelphia, Pennsylvania 19122, USA}

\author{E.A.~Chudakov}
\affiliation{Thomas Jefferson National Accelerator Facility, Newport News, 
Virginia 23606, USA}

\author{E.~Cisbani}

\author{S.~Colilli}                                                   
\affiliation{Istituto Nazionale di Fisica Nucleare, Sezione di Roma1, gruppo
collegato  Sanit\`a, and Istituto Superiore di Sanit\'a, I-00161 Roma, Italy}

\author{L.~Coman}
\affiliation{Florida International University, Miami, Florida 33199, USA}

\author{B.J.~Craver}
\affiliation{University of Virginia, Charlottesville, Virginia 22904, USA}

\author{G.~De~Cataldo}
\affiliation{Istituto Nazionale di Fisica Nucleare, Sezione di Bari and 
University of Bari, I-70126 Bari, Italy} 

\author{C.W.~de~Jager}
\affiliation{Thomas Jefferson National Accelerator Facility, Newport News, 
Virginia 23606, USA}

\author{R.~De~Leo}
\affiliation{Istituto Nazionale di Fisica Nucleare, Sezione di Bari and 
University of Bari, I-70126 Bari, Italy} 

\author{A.P.~Deur}
\affiliation{University of Virginia, Charlottesville, Virginia 22904, USA}

\author{C.~Ferdi}
\affiliation{Universit\'{e} Blaise Pascal/IN2P3, F-63177 Aubi\`{e}re, France}

\author{R.J.~Feuerbach}
\affiliation{Thomas Jefferson National Accelerator Facility, Newport News, 
Virginia 23606, USA}

\author{E.~Folts}
\affiliation{Thomas Jefferson National Accelerator Facility, Newport News, 
Virginia 23606, USA}

\author{R.~Fratoni}

\author{S.~Frullani}                                                   
 
\author{F.~Garibaldi}
\affiliation{Istituto Nazionale di Fisica Nucleare, Sezione di Roma1, gruppo 
collegato  Sanit\`a, and Istituto Superiore di Sanit\'a, I-00161 Roma, Italy}

\author{O.~Gayou}
\affiliation{Massachussets Institute of Technology, Cambridge, Massachusetts 
02139, USA}

\author{F.~Giulani}           
\affiliation{Istituto Nazionale di Fisica Nucleare, Sezione di Roma1, gruppo 
collegato  Sanit\`a, and Istituto Superiore di Sanit\'a, I-00161 Roma, Italy}
 
 \author{J.~Gomez}
 \affiliation{Thomas Jefferson National Accelerator Facility, Newport News, 
Virginia 23606, USA}

\author{M.~Gricia} 
\affiliation{Istituto Nazionale di Fisica Nucleare, Sezione di Roma1, gruppo 
collegato  Sanit\`a, and Istituto Superiore di Sanit\'a, I-00161 Roma, Italy}

\author{J.O.~Hansen}
\affiliation{Thomas Jefferson National Accelerator Facility, Newport News, 
Virginia 23606, USA}

\author{D.~Hayes}
\affiliation{Old Dominion University, Norfolk, Virginia 23508, USA} 

\author{D.W.~Higinbotham}
\affiliation{Thomas Jefferson National Accelerator Facility, Newport News, 
Virginia 23606, USA} 

\author{T.K.~Holmstrom}
\affiliation{College of William and Mary, Williamsburg, Virginia 23187, USA}

\author{C.E.~Hyde}
\affiliation{Old Dominion University, Norfolk, Virginia 23508, USA} 
\affiliation{Universit\'{e} Blaise Pascal/IN2P3, F-63177 Aubi\`{e}re, France}

\author{H.F.~Ibrahim}
\affiliation{Old Dominion University, Norfolk, Virginia 23508, USA} 

\author{X.~Jiang}
\affiliation{Rutgers, The State University of New Jersey, Piscataway, 
New Jersey 08855, USA}  

\author{L.J.~Kaufman}
\affiliation{University of Massachussets Amherst, Amherst,  Massachusetts 
01003, USA}

\author{K.~Kino}
\affiliation{Research Center for Nuclear Physics, Osaka 
University, Ibaraki, Osaka 567-0047, Japan}

\author{B.~Kross}
\affiliation{Thomas Jefferson National Accelerator Facility, Newport News, 
Virginia 23606, USA}

\author{L.~Lagamba}
\affiliation{Istituto Nazionale di Fisica Nucleare, Sezione di Bari and 
University of Bari, I-70126 Bari, Italy} 

\author{J.J.~LeRose}
\affiliation{Thomas Jefferson National Accelerator Facility, Newport News, 
Virginia 23606, USA}

\author{R.A.~Lindgren}
\affiliation{University of Virginia, Charlottesville, Virginia 22904, USA}

\author{M.~Lucentini}  
\affiliation{Istituto Nazionale di Fisica Nucleare, Sezione di Roma1, gruppo 
collegato  Sanit\`a, and Istituto Superiore di Sanit\'a, I-00161 Roma, Italy}

\author{D.J.~Margaziotis}
\affiliation{ California State University, Los Angeles, Los Angeles, 
California 90032, USA}

\author{P.~Markowitz}
\affiliation{Florida International University, Miami, Florida 33199, USA}

\author{S.~Marrone}
\affiliation{Istituto Nazionale di Fisica Nucleare, Sezione di Bari and 
University of Bari, I-70126 Bari, Italy} 

\author{Z.E.~Meziani}
\affiliation{Temple University, Philadelphia, Pennsylvania 19122, USA}

\author{K.~McCormick}
\affiliation{Rutgers, The State University of New Jersey, Piscataway, 
New Jersey 08855, USA}  

\author{R.W.~Michaels}
\affiliation{Thomas Jefferson National Accelerator Facility, Newport News, 
Virginia 23606, USA}

\author{D.J.~Millener}
\affiliation{Brookhaven National Laboratory, Upton, New York 11973, USA} 

\author{T.~Miyoshi}
\affiliation{Tohoku University, Sendai, 980-8578, Japan}

\author{B.~Moffit}
\affiliation{College of William and Mary, Williamsburg, Virginia 23187, USA}

\author{P.A.~Monaghan}
\affiliation{Massachussets Institute of Technology, Cambridge, Massachusetts 
02139, USA}

\author{M.~Moteabbed}
\affiliation{Florida International University, Miami, Florida 33199, USA}

\author{C.~Mu\~noz~Camacho}
\affiliation{CEA Saclay, DAPNIA/SPhN, F-91191 Gif-sur-Yvette, France}

\author{S.~Nanda}
\affiliation{Thomas Jefferson National Accelerator Facility, Newport News, 
Virginia 23606, USA}

\author{E.~Nappi}
\affiliation{Istituto Nazionale di Fisica Nucleare, Sezione di Bari and 
University of Bari, I-70126 Bari, Italy} 

\author{V.V.~Nelyubin}
\affiliation{University of Virginia, Charlottesville, Virginia 22904, USA}

\author{B.E.~Norum}
\affiliation{University of Virginia, Charlottesville, Virginia 22904, USA}

\author{Y.~Okasyasu}
\affiliation{Tohoku University, Sendai, 980-8578, Japan}

\author{K.D.~Paschke}
\affiliation{University of Massachussets Amherst, Amherst,  Massachusetts 
01003, USA}

\author{C.F.~Perdrisat}
\affiliation{College of William and Mary, Williamsburg, Virginia 23187, USA}

\author{E.~Piasetzky}
\affiliation{School of Physics and Astronomy, Sackler Faculty of Exact Science, 
Tel Aviv University, Tel Aviv 69978, Israel}

\author{V.A.~Punjabi}
\affiliation{Norfolk State University, Norfolk, Virginia 23504, USA}

\author{Y.~Qiang}
\affiliation{Massachussets Institute of Technology, Cambridge, Massachusetts 
02139, USA}

\author{B.~Raue}
\affiliation{Florida International University, Miami, Florida 33199, USA}

\author{P.E.~Reimer}
\affiliation{Argonne National Laboratory, Argonne, Illinois 60439, USA}

\author{J.~Reinhold}
\affiliation{Florida International University, Miami, Florida 33199, USA}

\author{B.~Reitz}
\affiliation{Thomas Jefferson National Accelerator Facility, Newport News, 
Virginia 23606, USA}

\author{R.E.~Roche}
\affiliation{Florida State University, Tallahassee, Florida 32306, USA}

\author{V.M.~Rodriguez}
\affiliation{University of Houston, Houston, Texas 77204, USA}

\author{A.~Saha}
\affiliation{Thomas Jefferson National Accelerator Facility, Newport News, 
Virginia 23606, USA}

\author{F.~Santavenere}
\affiliation{Istituto Nazionale di Fisica Nucleare, Sezione di Roma1, gruppo 
collegato  Sanit\`a, and Istituto Superiore di Sanit\'a, I-00161 Roma, Italy}
 
\author{A.J.~Sarty}
\affiliation{St. Mary's University, Halifax, Nova Scotia, Canada}

\author{J.~Segal}
\affiliation{Thomas Jefferson National Accelerator Facility, Newport News, 
Virginia 23606, USA}

\author{A.~Shahinyan}
\affiliation{Yerevan Physics Institute, Yerevan, Armenia}

\author{J.~Singh}
\affiliation{University of Virginia, Charlottesville, Virginia 22904, USA}

\author{S.~\v{S}irca}
\affiliation{Dept. of Physics, University of Ljubljana, Slovenia}

\author{R.~Snyder}
\affiliation{University of Virginia, Charlottesville, Virginia 22904, USA}

\author{P.H.~Solvignon}
\affiliation{Temple University, Philadelphia, Pennsylvania 19122, USA}

\author{M.~Sotona}
\affiliation{Nuclear Physics Institute, \v{R}e\v{z} near Prague, Czech
Republic}

\author{R.~Subedi}
\affiliation{Kent State University, Kent, Ohio 44242, USA} 

\author{V.A.~Sulkosky}
\affiliation{College of William and Mary, Williamsburg, Virginia 23187, USA}

\author{T.~Suzuki}
\affiliation{Tohoku University, Sendai, 980-8578, Japan}

\author{H.~Ueno}
\affiliation{Yamagata University, Yamagata 990-8560, Japan}

\author{P.E.~Ulmer}
\affiliation{Old Dominion University, Norfolk, Virginia 23508, USA} 

\author{G.M.~Urciuoli}
\affiliation{Istituto Nazionale di Fisica Nucleare, Sezione di Roma1, 
Piazza A. Moro, Rome, Italy}

\author{P.~Veneroni}
\affiliation{Istituto Nazionale di Fisica Nucleare, Sezione di Roma1, gruppo
collegato  Sanit\`a, and Istituto Superiore di Sanit\'a, I-00161 Roma, Italy}
 
\author{E.~Voutier}
\affiliation{LPSC, Universit\'e Joseph Fourier, CNRS/IN2P3, INPG, F-38026 Grenoble, France} 
 
\author{B.B.~Wojtsekhowski}
\affiliation{Thomas Jefferson National Accelerator Facility, Newport News, 
Virginia 23606, USA}

\author{X.~Zheng}                                                        
\affiliation{Argonne National Laboratory, Argonne, Illinois 60439, USA}

\author{C.~Zorn}
\affiliation{Thomas Jefferson National Accelerator Facility, Newport News, 
Virginia 23606, USA}

\collaboration{Jefferson Lab Hall A Collaboration}
\noaffiliation

\date{\today}

\begin{abstract}
An experiment measuring electroproduction of hypernuclei has been performed
in Hall A at Jefferson Lab  on a  $^{12}$C target.  In order to
increase counting rates and provide unambiguous kaon identification
 two superconducting septum magnets and a 
Ring Imaging CHerenkov detector (RICH) were added to the Hall A standard 
equipment. 
An unprecedented energy 
resolution of less than 700 keV FWHM has been achieved.
Thus, the observed \lam{12}{B} spectrum shows for the first time identifiable 
strength in the core-excited region between the ground-state {\it s}-wave 
$\Lambda$ peak and the 11 MeV  {\it p}-wave $\Lambda$ peak.

\end{abstract}


\maketitle

 Hypernuclei, long-lived baryonic systems with strangeness $\neq 0$,
provide us with a variety of nuclear phenomena. For example, a
$\Lambda$ hyperon (baryon with strangeness $\mathrm{S}\!=\!-1$ and mean 
lifetime $\tau\sim 10^{-10}$ s) can be placed deep inside the nucleus
as an impurity providing a sensitive probe of the nuclear interior.
The $\Lambda$ couples weakly to nuclear core states and the
$V_{N\Lambda}$ residual interaction removes the degeneracy of the
multiplets. In the case of a $\Lambda$ in an $s$ orbit, the
resulting doublet is split by the spin-dependent components of the 
interaction. 
The doublet spacings range from a few up to several hundred keV.
Since very limited information can be obtained from
elementary hyperon-nucleon scattering, hypernuclei are unique
laboratories for studying the $\Lambda$N interaction. 

In the past, hypernuclear spectroscopy has been carried out with
limited resolution by means of hadronic reactions, such as the
strangeness-exchange $^AZ(K^-,\pi^-)$\lamb{A}{Z}
and associated-production  $^AZ(\pi^+,K^+)$\lamb{A}{Z} reactions.
More recently, $\gamma$-ray spectroscopy has been used to measure
hypernuclear transition energies. Here, the few-keV energy resolution 
has allowed precise level assignments and the measurement of doublet
spacings~\cite{hashtam} but the method is limited to the bound region below
particle emission thresholds and to bound levels reached following particle emission.
 
The experimental knowledge can be enhanced using electroproduction
of strangeness that is characterized by a large 3-momentum transfer to the hypernucleus
($\gtrsim$ 250 MeV/c), 
a large angular momentum transfer $\Delta J$, and strong spin-flip 
terms, even at zero kaon production angle.  
Moreover, the $K^+ \Lambda$ pair production occurs on a proton in contrast
to a neutron in $(K^-,\pi^-)$ or $(\pi^+,K^+)$ reactions 
making possible the study of different hypernuclei and  charge-dependent
effects from a comparison of mirror hypernuclei.

The E94-107 experiment in Hall A at Jefferson Lab~\cite{proposal} 
(JLab) started a systematic study of high-resolution hypernuclear spectroscopy 
on $p$-shell targets, specifically  $^{9}$Be,  $ ^{12}$C, and $ ^{16}$O.  
The results on $ ^{12}$C are presented in this paper.

 $^{12}$C targets have been extensively used in hypernuclear studies
with the $(K^-,\pi^-)$, $(\pi^+, K^+)$ and $(K^-_\textrm{stop},\pi^-)$
reactions that are dominated by non-spin-flip contributions. 
In the early  experiments, only two peaks, attributed to the
$\Lambda$ in $s$ or $p$ orbits coupled to the $^{11}$C ground
state, were evident~\cite{hashtam}. The first evidence of structure 
between the main peaks came from $(\pi^+, K^+)$ studies with the SKS 
spectrometer at KEK (E140a, E336, and E369)~\cite{hashtam}, with the best 
resolution of 1.45 MeV in KEK E369~\cite{KEK369}. 
Recently, in the stopped $K^-$ experiment of the FINUDA 
collaboration~\cite{FINUDA}, further evidence for structure in this
region has been observed. 
The first electroproduction experiment~\cite{Miyoshi} performed on
a $^{12}$C target at JLab in Hall C had limited statistics but proved that
the electroproduction process can be used to study
hypernuclear spectra with sub-MeV energy resolution and
measured cross sections.

Hall A at JLab is well suited to perform $(e,e'K^+)$ experiments. 
Scattered electrons can be detected in the High Resolution Spectrometer
(HRS) electron arm while coincident kaons are detected in the HRS hadron 
arm~\cite{nimhalla}. The disadvantage of smaller electromagnetic cross 
sections is partially compensated for by the high current, high duty cycle, 
and high energy resolution capabilities of the beam at Jefferson Lab.
In the present experiment, a 100 mg/cm$^2$ $^{12}$C target 
was used with an electron beam current of 100 $\mu$A.

 The strong inverse dependence of the cross section on $Q^2$, squared 
virtual photon 4-momentum transfer, 
calls for measurements at low $Q^2$. 
To maximize the cross section,
the electron scattering angle must be minimized, 
subject to avoiding the increasing background
from processes at very forward electron angles.
To minimize the momentum transferred to the hypernucleus, and maximize the cross section,
a detection angle $\theta_K$ for the K$^+$ must be chosen near the virtual photon direction.
The high beam energy results
in a relatively high momentum for the kaon, as required to keep
a reasonable survival fraction in the spectrometer (25\,m flight path).
So, kinematics were set to particle detection at  $6^\circ$ for both 
electrons and kaons, incident beam energy of 3.77 GeV, scattered electron momentum 
of 1.56 GeV/c, and kaon momentum of 1.96 GeV/c. 

 In order to allow experiments at forward angles smaller than 
the  HRS's minimum angle (12.5$^\circ$), a superconducting septum magnet 
was added to each HRS. Particles at scattering angles of 
$6^\circ$ are deflected by the septum magnets into the HRS.
This new spectrometer configuration (Septum + HRS) provides a general
purpose device that extends the HRS features to small scattering angles  
while preserving the spectrometer optical performance~\cite{septum}.
The energy resolution depends on the momentum 
resolution  of the HRS spectrometers, on the straggling
and energy loss in the target, and on the beam energy spread. 
A momentum resolution of the system (HRS's + septum magnets)
of $\Delta p / p = 10^{-4}$ (FWHM) and a beam energy spread as 
small as $6\times 10^{-5}$ (FWHM) are necessary to be able to achieve
an excitation-energy resolution of about 500 keV.
With a  dedicated effort the Accelerator staff were able to address the
beam quality  requirements and to set up new devices for continuous beam energy
spread monitoring.

The high background level demands a very efficient 
PID system with unambiguous kaon identification. 
The standard PID system in the hadron arm is composed of two aerogel threshold
Cherenkov counters~\cite{lagamba,nimhalla} ($n_1 = 1.015$,  $n_2 = 1.055$). 
Charged pions (protons) with momenta around 2 GeV/c are above (below) the
Cherenkov light emission threshold. Kaons emit Cherenkov light only in the $n_2 = 1.055$
detector. Hence, a combination of the 
signals from the two counters should distinguish among the three species 
of hadrons. However, due to
inefficiencies and delta-ray production, 
the identification of kaons has contamination from pions 
and protons. This has driven the design, construction, and installation 
of a Ring Imaging CHerenkov (RICH) detector, conceptually identical to
the ALICE HMPID design~\cite{AliceRichTDR}, in the hadron HRS 
detector package.  It uses a proximity focusing 
geometry, a CsI  photocathode, and a 15 mm thick liquid perfluorohexane 
radiator. A detailed description of the layout and the performance of the 
RICH detector is given in~\cite{rich2004,pylosfg,pylosfc}.
In the electron arm, the Gas Cherenkov counters give pion rejection  
ratios up to $10^{3}$. The remaining backgound (due to knock-on electrons)
is reduced by a further 2 orders of magnitude by the lead glass pre-shower 
and shower counters, giving a total pion rejection ratio of $10^{5}$. 

\begin{figure}[t]
 \centering
\includegraphics[angle=0, width=8.6cm]{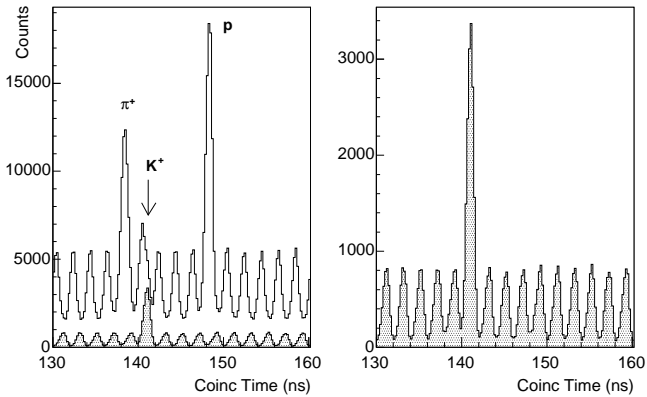}
\caption{Hadron plus electron arm coincidence time spectra. 
In the left panel, the unfilled histogram is obtained by selecting kaons with 
only the threshold aerogel Cherenkov detectors. The filled histogram 
(expanded in the right panel) includes the RICH kaon selection. 
The remaining contamination is due to accidental $(e,e')\otimes (e,K^+)$ 
coincidences.}
\label{ctkaon}
\end{figure}

\begin{figure}[t]
 \centering
 \includegraphics[width=8.6cm]{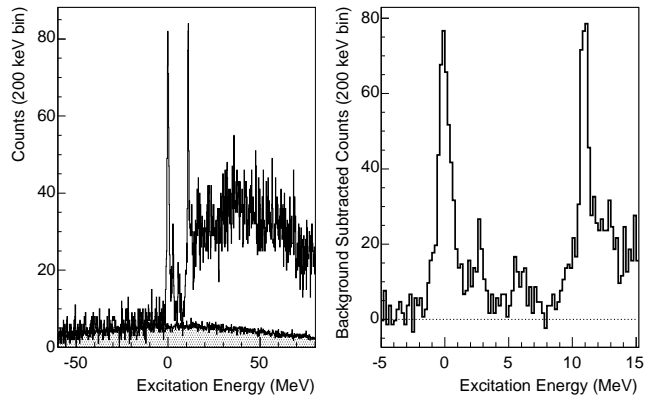}
\caption{The \lam{12}{B} excitation-energy spectrum obtained after 
kaon selection with aerogel detectors and RICH. The electron-kaon random 
coincidence contribution evaluated in a large timing window is superimposed 
on the spectrum in the left panel. The right panel shows the spectrum after 
this background has been subtracted.}
\label{carbon1}
\end{figure}

The essential role of the RICH in identifying kaons is shown 
in Fig.~\ref{ctkaon}, where the unfilled, without RICH,
should be compared to the filled spectrum, with RICH. 
All contributions from pions and protons completely vanish with the RICH. 

\begin{table*}[t]
\caption{\label{table} Levels and cross sections obtained by fitting the 
$^{12}$C$(e,e'K^+) ^{12}_{\Lambda}$B spectrum compared with theoretical 
predictions. In column 6, $p_\Lambda$ for the $1^+$ states indicates 
strongly mixed $p_{1/2}$ and $p_{3/2}$ configurations.}
\begin{ruledtabular},
\begin{tabular}{cccc|cccc}
 & \multicolumn{2}{c}{Experimental data} &  & & \multicolumn{2}{c}{Theoretical prediction}  \\
\hline
Position                &         Width                         &    SNR      &  Cross Section                                            & $E_x$  &     Main structure                                                                     & $J^\pi$        & Cross section \\
(MeV) & (FWHM, MeV) & &(nb/sr$^2$/GeV)  &  (MeV) & &  & (nb/sr$^2$/GeV) \\        
\hline 
$0.0\pm 0.03$    &          $1.15 \pm 0.18$     &     19.7      &      $4.48 \pm 0.29 (st) \pm 0.63 (sys)$    &   0.0      &     $^{11}$B$(\frac{3}{2}^-; g.s.) \otimes s_{1/2\Lambda}$      &      1$^-$     &   1.02    \\
                              &                                            &                   &                                                                         & 0.14     &                                         \hspace{1cm}    ''                                      &       2$^-$    &   3.66 \\
\hline
$2.65\pm 0.10$ &           $0.95 \pm 0.43$    &      7.0       &      $0.75\pm 0.16 (st) \pm 0.15 (sys)$      &  2.67    &    $^{11}$B$(\frac{1}{2}^-; 2.12) \otimes s_{1/2\Lambda}$     &       1$^-$    &  1.54 \\
\hline
$5.92\pm 0.13$ &           $1.13 \pm 0.29$    &      5.3       &       $0.45 \pm 0.13 (st) \pm 0.09 (sys)$     &  5.74    &    $^{11}$B$(\frac{3}{2}^-; 5.02) \otimes s_{1/2\Lambda}$     &       2$^-$   &    0.58 \\
                             &                                             &                  &                                                                          &   5.85    &                                             \hspace{1cm}    ''                                &       1$^-$   &    0.18 \\
\hline
$9.54\pm 0.16$ &            $0.93 \pm 0.46$   &      4.4       &        $0.63 \pm 0.20 (st) \pm 0.13 (sys) $    &    --       & --                                                                                                        & --                &         -- \\
\hline
$10.93\pm 0.03$ &          $0.67\pm 0.15$    &   20.0       &         $3.42 \pm 0.50 (st) \pm 0.55 (sys)$     & 10.48  &     $^{11}$B$(\frac{3}{2}^-;g.s.)\otimes p_{3/2\Lambda}$        &  2$^+$ & 0.24   \\
                               &                                           &                  &                                                                            &  10.52 &     \hspace{0.8cm}    ''  \hspace{1cm}     $p_{\Lambda}$    &  1$^+$ & 0.12 \\
                               &                                           &                  &                                                                            &  10.98 &     \hspace{1cm}    ''  \hspace{0.8cm} $p_{1/2\Lambda}$         &  2$^+$ & 1.43 \\
                               &                                           &                  &                                                                            &  11.05 &     \hspace{1cm}    ''  \hspace{0.8cm} $p_{3/2\Lambda}$         &  3$^+$ & 2.19 \\
\hline
$12.36\pm 0.13 $ &         $ 1.58\pm 0.29$   &     7.3      &         $1.19 \pm 0.36 (st) \pm 0.35 (sys) $     &  12.95 &      $^{11}$B$(\frac{1}{2}^-; 2.12) \otimes p_{3/2\Lambda}$     &   2$^+$    &  0.91 \\
                                &                                          &                 &                                                                             &  13.05 &        \hspace{0.8cm}    ''  \hspace{1cm} $p_{\Lambda}$             &   1$^+$   & 0.27 \\
\end{tabular}
\end{ruledtabular}
\end{table*}

\begin{figure}[h]
 \centering
\includegraphics[width=9.0cm]{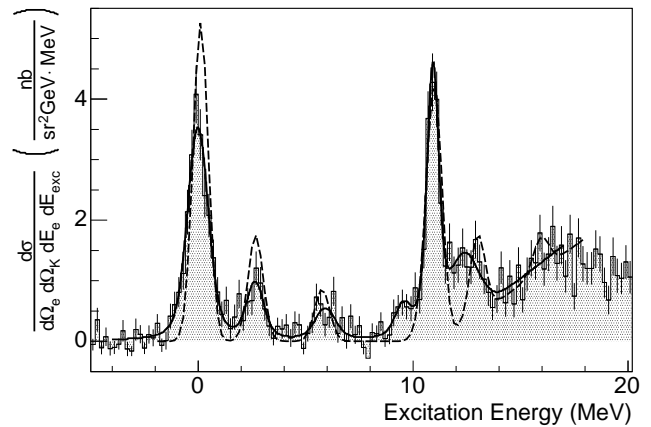}
\caption{The \lam{12}{B} excitation-energy spectrum. The best fit (solid curve)
and a theoretical prediction (dashed curve) are superimposed on the 
data. See text for details.}
\label{carbon2}
\end{figure}

In Fig.~\ref{carbon1}, the excitation energy spectrum of \lam{12}{B} 
is shown for the full range of energy acceptance. The filled
histogram shows the low level of $(e,e')$$\otimes$$(e,K^+)$ random 
coincidence background. 
Figure~\ref{carbon2} shows the six-fold differential cross section 
expressed in nb/(sr$^2$ GeV MeV). 
The background has been evaluated by
fitting the data obtained for random coincidences in a large
timing window. 
No residual background in the negative range
of $E_{x}$ is present after subtraction. 
The origin of the excitation energy scale
has been set to the peak value of the ground-state (g.s.) level
(the uncertainty of the absolute scale being about 0.5 MeV).

The first step in fitting the data was to use
a peak search algorithm \cite{peaksearch} to identify six regions with
an excess of counts above background at the 90$\%$ confidence level. 
When fitted with Gaussian functions, individual peaks show non-gaussian contributions, mainly
in the tails due to the radiative effects.
Voigt functions \cite{voigt},
convolutions of gaussian with Lorentzian functions, are often used in
spectroscopy to better fit the data. 
The main idea in fitting the data is to minimize the number of assumptions. 
Since each peak might contain a more complex (unresolved) structure, 
each Voigt function allows the width, location and height to vary independently.
The best fit is thus determined by minimizing  $\chi ^2$/n.d.f.
with respect to the positions, widths, and amplitudes of 6 Voigt functions and 
parameters of the quasi-free region modeled by a quadratic form.
The results of the parameters obtained in the fit with $\chi ^2$/n.d.f.=1.16, 
together with the statistical significance of the assigned 
levels (in terms of $SNR = \textrm{Signal}/\sqrt{(\textrm{Signal} + 
\textrm{BckGrnd})}$) and cross sections, as obtained after a radiative 
unfolding procedure, are given in Table~\ref{table}.
 
The narrowest width of $ 670 \pm 150$ keV has been measured for
the peak at $E_x=10.93$ MeV, 
indicating that the experimental excitation energy resolution is 
at least compatible with this value.

On the other hand, the width for the g.s. peak, around 
$E_x = 0.0$ MeV, is $1150 \pm 180$ keV, wider than 670 keV, although, 
within errors, not in sharp statistical disagreement. 

Such a larger width might suggest a more complex structure underlying
the g.s. peak. The possibility of an unresolved doublet
has been explored by fitting with two Voigt functions, constraining
their widths to be the same as the peak at $E_x=10.93$ MeV.
The result is that the separation between the two Voigt shapes is
about 650 keV, larger than the 140 keV predicted by the theory.
However, with the present sample of data, there is not enough statistical
significance to favor such a result over a simple statistical fluctuation
of the widths. This situation could be clarified either with a larger
sample of data or by an improvement in the experimental resolution.

 Due to the very low level of background, states with an $s_\Lambda$ coupled
to excited $^{11}$B core states are clearly observed   
between the g.s. and the level at 10.93 MeV with signal to noise 
ratios (SNR) larger than 5. The positions of these levels can be determined 
with uncertainties less than 150 keV. 
Cross sections are determined at the level of 15--20\%.

 In Fig.~\ref{carbon2}, the measured electroproduction cross sections for 
hypernuclear states are also compared with a model (dashed line), which  
shows very good overall agreement with the data
without any normalization factor. The theoretical cross sections  were
obtained in the framework of the Distorted Wave Impulse Approximation 
(DWIA)~\cite{modMS} using the Saclay-Lyon (SLA) model~\cite{SLA} for the 
elementary $p(e,e'K^{+})\Lambda$ reaction.
Shell-model wave functions for $^{11}$B and \lam{12}{B} were 
obtained using fitted $p$-shell interactions and a parametrization 
of the $\Lambda N$ interaction  that fits the precise
$\gamma$-ray spectra of \lamb{7}{Li}~\cite{modJM}. The results are
compared with experiment in Table~\ref{table}.

  The large g.s. peak and another strong peak at $\sim$ 10.93 MeV 
correspond to the substitution of a $p$-shell proton 
by a $\Lambda$ in $s$ and $p$ states, respectively, coupled dominantly to the 
$3/2^-$ g.s. of $^{11}$B. Two peaks at 
$E_x = 2.65$ MeV and $E_x = 5.92$ MeV are also evident. Theoretically
(see Table~\ref{table}), they should be
due to the $1^-$ member of the $0^-,\,1^-$ doublet based on the 
$1/2^-$ state of $^{11}$B at $E_x=2.125$ MeV and to the $1^{-},\,2^{-}$ 
doublet based on the $3/2^{-}$ level of $^{11}$B at $E_x = 5.020$ MeV, 
with the dominant contribution coming from the $2^{-}$ state. The raising
of the doublet energies relative to the unperturbed core energies is mainly
due to the $\vec{l}_{N\Lambda}\cdot \vec{s}_N$ component of the effective
$\Lambda N$ interaction~\cite{millhyp00}. 
The energies and cross sections for the three $s_\Lambda$ peaks
are well reproduced by the theoretical predictions for the five
$s_\Lambda$ states.

 Because the $\Lambda$ spin-orbit interaction is weak, the calculation 
predicts essentially degenerate $2^{+}$ and $3^{+}$ states from 
$^{11}$B$(3/2^-)\otimes p_\Lambda$  coupling. Table~\ref{table} shows
that these two states dominate in the main peak observed  close to 11 MeV. 
There is clearly strength on either side of this peak, accounted for in 
the fit by peaks at 9.54 MeV (though at the limit of statistical significance) 
and at 12.36 MeV. The theoretical $p_\Lambda$ strength
based on the ground and first-excited states of $^{11}$B accounts 
for 98\% of the observed strength. There are six known positive-parity
states between 9 and 12 MeV in $^{11}$B \cite{ajzenberg} to which an 
$s_\Lambda$ can couple to form $2^+$ or $3^+$ hypernuclear states. 
Based on existing $(e,e'p)$ data, these states are expected to be only 
weakly excited and an admixture with the nearby $p_\Lambda$ states
is required for them to be excited as strongly as observed in the
present data. From their shell-model structure, the 9.88 MeV $3/2^+$ 
and 11.60 MeV $5/2^+$ states of $^{11}$B should be the most important 
but a full $1h\omega$ shell-model calculation is needed
to investigate this problem.

 In summary, a high-quality, background-free \lam{12}{B} hypernuclear 
spectrum with unprecedented energy resolution ($\sim 670$ keV) has 
been obtained. The new experimental devices have proven to be very 
effective. In particular, septum magnets had no adverse affect on
the HRS optics and the RICH played an essential role in the
unambiguous identification of kaons.

The measured cross section for the g.s. doublet is in 
very good agreement with the value of 4.68 nb/sr$^2$/GeV predicted
using the SLA model. The $s_\Lambda$ part of the spectrum is well 
reproduced by the theory and a very good agreement is also obtained for the 
other levels. For the first time a measurable strength with good 
energy resolution has been observed  in the core-excited part of 
the spectrum. This is helped by the fact that the spin-spin 
interaction enhances these states with respect to the weak-coupling 
limit. The distribution of strength within several MeV on either 
side of the strong $p_\Lambda$ peak should stimulate theoretical 
work to better understand the $p_\Lambda$ region.

We acknowledge the Jefferson Lab physics and accelerator 
Division staff for the outstanding efforts that made
this work possible. 
This work was supported by U.S. DOE contract 
DE-AC05-84ER40150, Mod. nr. 175,
under which the Southeastern Universities Research Association (SURA)
operates the Thomas Jefferson National Accelerator Facility,
 by the Italian Istituto Nazionale di Fisica 
Nucleare and  by the Grant Agency of the Czech Republic under grant No. 
202/05/2142, and by the U.S. DOE under contracts, DE-AC02-06CH11357,
DE-FG02-99ER41110, and DE-AC02-98-CH10886.


\end{document}